\documentclass[aps,prl,onecolumn,showpacs]{revtex4}
\usepackage{amssymb}

\usepackage{graphicx}

\begin{document}

\title{Note on possibility of proximity induced spontaneous currents in superconductor/normal
metal heterostructures}
\author{Anatoly F. Volkov and Ilya M. Eremin}
\affiliation{Institut f\"ur Theoretische Physik III, Ruhr-Universit\"at Bochum, D-44801
Bochum, Germany}
\date{\today }

\begin{abstract}
We analyse the possibility of the appearance of spontaneous currents in
proximated superconducting/normal metal (S/N) heterostructure when Cooper
pairs penetrate into the normal metal from the superconductor. In
particular, we calculate the free energy of the S/N structure. We show that
whereas the free energy of the N film $F_{N}$ in the presence of the
proximity effect increases compared to the normal state, the total free
energy, which includes the boundary term $F_{B}$, decreases. The condensate
current decreases $F_{N}$, but increases the total free energy making the
current-carrying state of the S/N system energetically unfavorable.
\end{abstract}

\maketitle
\affiliation{Theoretische Physik III,\\
Ruhr-Universit\"{a}t Bochum, D-44780 Bochum, Germany}
\date{\today }

\textit{Introduction}. Penetration of Cooper pairs into the normal metal (N)
in superconductor/normal metal (S/N) heterostructures, provided the
interface transparency is not too small, is a well-known effect 
\cite{deGennesRMP-Book65,Lambert98,TinkhamBook96,DeutscherRMP05}. This so-called
proximity effect (PE) is related to the Andreev reflections of electrons at
the interface of the S/N bilayer \cite{AndreevJETP64}. In particular, the
depth of Cooper pairs penetration into the N in the diffusive case is equal
to $\xi _{N}\cong \sqrt{D_{N}/2\pi T}$ where $D_{N}=vl/3$ is the diffusion coefficient, and becomes smaller if the condensate moves.
The proximity effect is utilized in various S/N/S Josephson junctions \cite{KulikBook72,LikharevRMP79,BaroneBook82} and other superconducting devices\cite{BirgePRB08,BelzigPRApp20,GiazottoPRB09,LinderRev15} as it leads to a
number of interesting physical phenomena. The most famous of these is the
advantageous Josephson coupling in S/N/S Josephson junctions with the N
layer being significantly thicker (up to a few microns) than the insulating
(I) barrier in tunnel Josephson junctions \cite{KulikBook72,LikharevRMP79,BaroneBook82}. Furthermore, in contrast to the
conventional tunnel S/I/S junction, the properties of the Josephson S/N/S
junctions can be modified by varying the characteristics of the normal metal
layer. For example, if there is an exchange field in the N metal, i. e. a
ferromagnetic metallic layer F is used, then the critical current $I_{c}$%
may even change sign \cite{GolubovRMP04,BuzdinRMP05,BVErmp05,EschrigRev11,LinderRev15,LinderBalRMP17},
yielding the so-called $\pi $-junctions. Note, the change of sign
in $I_{c}$ may also be achieved in conventional S/N/S
multi-terminal Josephson junctions if the electric potential of the normal
metal N is shifted with respect to the S counterparts \cite%
{VolkovPRL95,ZaikinPRL98,YipPRB98,KlapwijkNature99,BirgePRB08}. More
recently, spectrum of Andreev bound states in S/N-multiterminal structures
with potentially non-trivial topological bands with Weyl points was also
investigated \cite{Nazarov16,NazarovPRB21}. 

Despite of these continuous research efforts in simple S/N systems
and their derivatives, outlined above, the origin of certain effects remains
mysterious. For example, an interesting paramagnetic re-entrant effect
(sometime called Mota effect) caused by spontaneous currents in S/N bilayer
was observed in Refs.\cite{Mota90,Mota94,Mueller99,MotaPRL00}. The authors
of Ref.\cite{BelzigBlatterPRL99} proposed an explanation in terms of a
repulsive interaction with a negative small coupling constant $\lambda _{N}$%
 i.e. assuming the normal metal may acquires a gap, $\Delta _{N}$, which sign is opposite to that in a superconductor, $\Delta $. However, the predicted paramagnetic response caused by spontaneous
currents turned out to be too small because of the smallness of the
superconducting order parameter in N $\Delta _{N}\sim \lambda _{N}$, and thus the origin of the Mota effect remains unclear \cite{MotaPRL00}.
Note, the paramagnetism and spontaneous currents may occur in S/F system\cite{BVEprb01,BuzdinPRL12,VBEprb19,BuzdinPRB19} or S/F/N structures \cite{HaltermanPRB14,MarychevPRB20}. However, there the origin of the
paramagnetic effect should be quite different from that in S/N structures as
in the former it is related to internal exchange fields existing in the
ferromagnet F and to the triplet Cooper pairs induced in the film F by the
PE \cite{GolubovRMP04,BuzdinRMP05,BVErmp05,EschrigRev11,LinderRev15,LinderBalRMP17}.
In S/N/S Josephson junctions in a non-equilibrium \cite{BelzigPRL20}
spontaneous currents arise when the Josephson current in S/N/S junctions
changes sign \cite{VolkovPRL95,ZaikinPRL98,YipPRB98,KlapwijkNature99,BirgePRB08} but this
situation then resembles the case of S/F/S junctions with a negative
Josephson current \cite{GolubovRMP04,BuzdinRMP05,BVErmp05,EschrigRev11,LinderRev15,LinderBalRMP17}.
Therefore, the situation of the S/N bilayer in equilibrium requires a
separate study.

In this paper, we consider a simple S/N bilayer heterostructure with a
superconducting coupling constant in the N layer equal to zero, i. e., $%
\lambda _{N}=0$\ and $\Delta _{N}=0$. We calculate the total free energy $%
F_{S/N}$\ of the system that consists of bulk terms $F_{S}$\ and $F_{N}$\ as
well as the boundary term $F_{B}$. Below the critical temperature $T_{c}$,
the energy $F_{S}$\ ($F_{N}$) decreases (increases), respectively. On the
contrary to $F_{N}$, the boundary term $F_{B}$\ decreases the total free
energy in such a way that the contribution of the terms $F_{N}+F_{B}$\ is
negative. The contribution $F_{S}+F_{B}$\ remains negative as it is in the
absence of the PE. The condensate current gives a positive contribution to
both terms $F_{N}+F_{B}$ and $F_{S}+F_{B}\,\ $making the current-carrying
state unfavorable.

\textit{Theory.} Frequently the analysis of the free energy ($F$) is
performed using the Ginzburg-Landau free energy expansion, assuming the
smallness of the order parameter $\Delta $. This approach is not applicable
to the considered heterostructure because the superconducting order
parameter $\Delta _{N}$ in the N film is assumed to be zero. On the other
hand, a part of electrons in N condense due to the PE and therefore the free
energy $F_{N}$ changes also in the superconducting state. Thus, in order to
calculate the variation $\delta F$, we need to find first the quasiclassic
matrix Green's functions $\hat{g}$ in the S and N regions using the boundary
conditions and to express the free energy in terms of the functions $\hat{g}$%
. We consider a simple case of diffusive S/N structure when the function $%
\hat{g}$ obeys the Usadel equation \cite{UsadelPRL70}. In particular, the
system under consideration is a bilayer which consists of S and N films with
thicknesses $d_{S,N}$, respectively as shown in Fig.1. The current is
assumed to flow along the interface in the $y$-direction. We integrate out
the phase $\chi (y)$ by making the transformation $\hat{g}_{n}=\hat{S}%
^{\dagger }\cdot \hat{g}\cdot \hat{S}$, where $\hat{S}=\exp [(iQy/2)\hat{\tau%
}_{3}]$. This means that the phase $\chi $ and the functions $\hat{g}_{n}$
after the transformation depend only on the $x$ coordinate and we drop the
subscript $"n"$ in what follows. We represent the matrix $\hat{g}$ in a
standard form $\hat{g}=\hat{\tau}_{3}\cos \theta +\hat{\tau}_{1}\sin \theta $%
, which is typically used in studying S/N structures \cite%
{VZKlap93,ZaikinRev99,FominovPRB01,VirtanenPRL04,LevchenkoPRB08,MaslovPRB14}
\begin{figure}[tbp]
	\includegraphics[width=0.2\columnwidth]{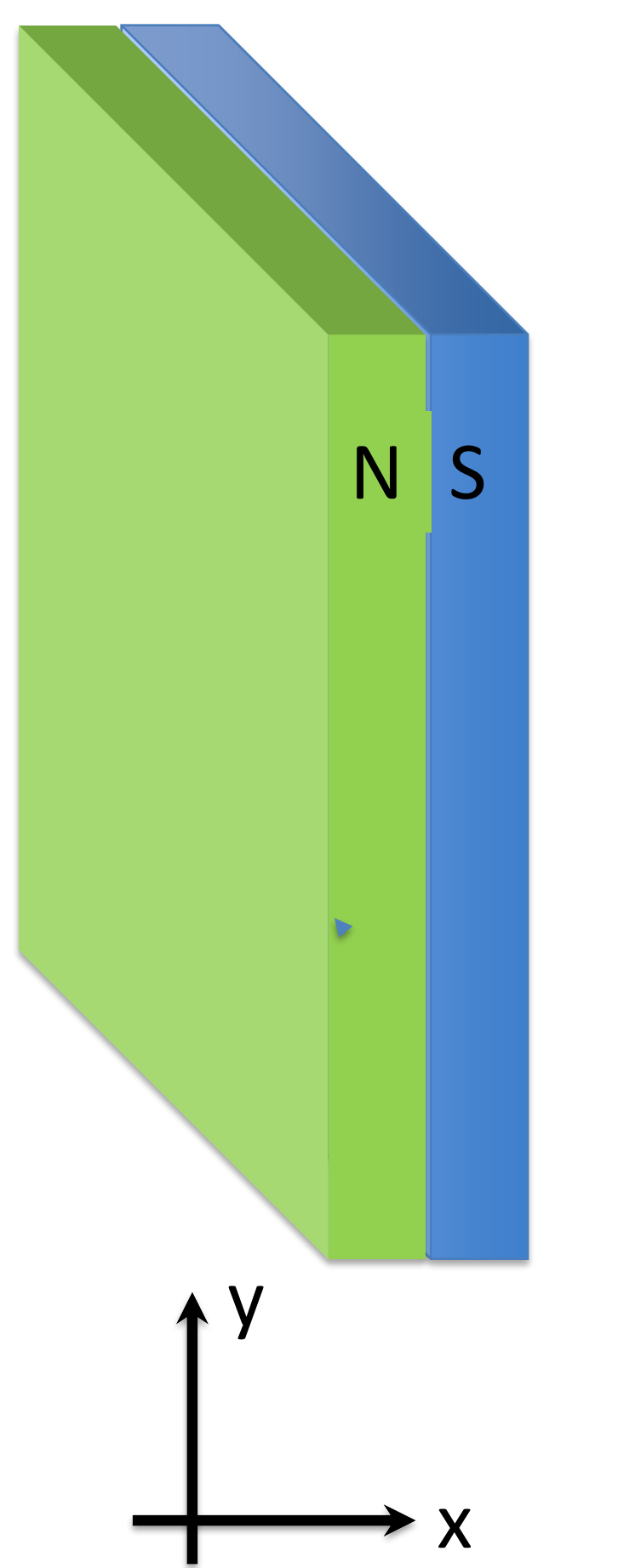} \vspace{-1mm} \vspace{-1mm}
	\caption{(Color online.) A schematic representation of the S/N bilayer
		structure.  }
	\label{S/N}
\end{figure}
so that the normalization condition $\hat{g}\cdot \hat{g}=\hat{1}$ is
automatically fulfilled. The function $\theta $ depends on $x$ and obeys the
Usadel equations in the S and N regions

\begin{eqnarray}
-D_{S}\partial _{xx}^{2}\theta _{S}+2\omega \sin \theta _{S}-2\Delta \cos
\theta _{S}+(D_{S}P_{S}^{2}/2)\sin (2\theta _{S}) &=&0\text{, S film }
\label{2} \\
-D_{N}\partial _{xx}^{2}\theta _{N}+2\omega \sin \theta
_{N}+(D_{N}P_{N}^{2}/2)\sin (2\theta _{N}) &=&0\text{, \ N film}  \label{2a}
\end{eqnarray}%
where $D_{S;N}$ are the diffusion coefficients in the S(N) films, $\omega $
is the Matsubara frequency, $\mathbf{P}=\mathbf{\nabla }\chi -2\pi \mathbf{A}%
/\Phi _{0}$ is the gauge-invariant condensate momentum and $\Phi _{0}=hc/2e$
is the magnetic flux quantum. The Usadel equations are complemented by the
standard Kurpiyanov-Lukichev boundary conditions for $\theta _{S,N}$\cite%
{KL88} at the interface
\begin{equation}
\partial _{x}\theta _{S(N)}=-\kappa _{B,S(N)}\sin (\theta _{S}-\theta
_{N})|_{x=0}\text{ }  \label{3}
\end{equation}%
where $\kappa _{B,S(N)}=2/R_{B}\sigma _{S(N)}$, $R_{B}$ is the S/N interface
resistance per unit area and $\sigma _{S,N}\,$are the conductivities in the
S and N films in the normal state. The order parameter $\Delta $, which is
non-zero in the S film, is determined by the self-consistency equation
\begin{equation}
\Delta =\lambda (2\pi T)\sum_{\omega \geqslant 0}\sin \theta _{S}(\omega )%
\text{ }  \label{4}
\end{equation}%
Note, Eq.(\ref{4}) and Eqs.(\ref{2}-\ref{2a}) are obtained by the variation
of the total free energy $F_{S}$ and $F_{N}$ with respect to $\Delta $ and $%
\theta $

\begin{eqnarray}
F_{S} &=&2\nu _{S}\int_{-d_{S}}^{0}dx\Bigg\{\frac{\Delta ^{2}}{2\lambda }%
+2\pi T\sum_{\omega \geqslant 0}\left[ \frac{D_{S}}{4}(\partial _{x}\theta
_{S})^{2}+\omega (1-\cos \theta _{S})-\Delta \sin \theta +\frac{D_{S}P^{2}}{8%
}\left( 1-\cos (2\theta _{S})\right) \right] \Bigg\}  \label{5} \\
\text{ }F_{N} &=&2\nu _{N}\int_{0}^{d_{N}}dx\Bigg\{2\pi T\sum_{\omega
\geqslant 0}\left[ \frac{D_{N}}{4}(\partial _{x}\theta _{N})^{2}+\omega
(1-\cos \theta _{N})+\frac{D_{N}P^{2}}{8}\left( 1-\cos (2\theta _{N})\right) %
\right] \Bigg\}  \label{5a}
\end{eqnarray}%
where $\nu $, $P$, $D$ are the density of states, momentum, and diffusion
coefficient in either S or N film, respectively. We set $\Delta _{N}$ equal
to zero since we assume that $\lambda _{N}=0$. The energy $F$ is counted
from its value in the normal state, \textit{i.e.} $\theta =0$. This
expression for $F_{N}$ can be also derived from a more general expression
for the free energy of a superconductor in the presence of an exchange field %
\cite{FuldePR65,SilaevPRB20}. We also note by passing that taking the
variation of the sum of the $F$ and the magnetic energy $(\mathbf{\nabla
\times A})^{2}/8\pi $ one obtains the London equation $\nabla ^{2}\mathbf{A}%
=(4\pi /c)\mathbf{j}$, where $\mathbf{j}=-(c/4\pi )\Lambda _{L}^{-2}\mathbf{P%
}$. Here, $\Lambda _{L}^{-2}=[2\sigma /(c^{2}\hbar )](2\pi T)\sum_{\omega
\geqslant 0}\sin ^{2}\theta (\omega ))$ is the inverse squared London
penetration depth. In order to take into account the boundary conditions (%
\ref{3}), we need \ to add the boundary term $F_{B}$ \cite%
{Altland00,Giazotto16} to $F_{S}+F_{N}$ so that the total functional $F$ is
given by

\begin{equation}
F=F_{S}+F_{N}+F_{B}\text{ }  \label{6}
\end{equation}

In the following we solve Eqs.(\ref{2}-\ref{2a}) for the functions $\theta
_{S,N}$ together with the self-consistency equation (\ref{4}) and find a
minimum of the free energy $F$ as a function of the condensate velocity $%
\mathbf{V=P}/m$. In a general case, this can be done only numerically. Here
we restrict the analysis with the simplest case of a weak proximity effect
when the Usadel equation for $\theta _{N}$\ref{2a} can be linearized and the
function $\theta _{S}$ is weakly perturbed by the PE. The latter assumption
is valid if the condition $\delta \theta _{S}\lesssim \xi _{S}/(R_{B}\sigma
_{S})\ll 1$ is fulfilled, where $\xi _{S}\cong \sqrt{D_{S}/2\Delta }$ is a
coherence length in S. Yet we do take into account a suppression of the
order parameter $\Delta $ by the condensate flow. In the case of small
suppression of $\Delta $, we find $\delta \Delta \cong
-(D_{S}P_{S}^{2}/2\Delta )\sum_{\omega \geqslant 0}(\omega ^{2}/\zeta
_{\omega }^{4})/\sum_{\omega \geqslant 0}\zeta _{\omega }^{-3}$, where $%
\zeta _{\omega }=\sqrt{\omega ^{2}+\Delta _{0}^{2}}$. At low temperatures ($%
T\ll \Delta )$ the gap variation is $\delta \Delta \cong -$ $%
D_{S}P_{S}^{2}/2 $. Note that a strong suppression of $\Delta $ by the
condensate flow was studied in Refs.\cite{FuldePR65,EstevePRL03}. In the
absence of the PE and the condensate flow, one has $\sin \theta _{S0}\equiv
f_{S}=\Delta _{0}/\zeta _{\omega }$ and $\cos \theta _{S0}=\omega /\zeta
_{\omega }$ with $\zeta _{\omega }=\sqrt{\omega ^{2}+\Delta _{0}^{2}}$. The
direct calculation of $F_{S0}$ gives a well known result $F_{S0}=-\nu
_{S}\Delta ^{2}d_{S}/2$ \cite{deGennesRMP-Book65}. The correction $\delta
F_{S0}$ caused by the condensate flow is $\delta F_{S0}=\nu
_{S}D_{S}P_{S}^{2}(2\pi T)\Delta \sum_{\omega \geqslant 0}\zeta _{\omega
}^{-2}$. Thus, the energy $F_{S}$ of the S film with a spontaneous current
is
\begin{equation}
F_{S}=-\nu _{S}d_{S}\frac{\Delta ^{2}}{2}\left[ 1-D_{S}P_{S}^{2}(2\pi
T)\sum_{\omega \geqslant 0}\zeta _{\omega }^{-2}\right]  \label{7}
\end{equation}%
and as expected the condensate flow reduces the condensation energy.

Next we evaluate the contribution to the free energy of the N film, $F_{N}$.
Linearized Eq.(\ref{2a}) has the form
\begin{equation}
-\partial _{xx}^{2}\theta _{N}+\kappa _{q}^{2}\theta _{N}=0\text{ }
\label{8}
\end{equation}%
with a solution

\begin{equation}
\theta _{N}(x)=\frac{\kappa _{B}}{\kappa _{q}}f_{S}\frac{\cosh (\kappa
_{q}(x-d_{N}))}{\sinh (\kappa _{q}d_{N})}\text{,}  \label{9}
\end{equation}%
where $\kappa _{q}=\sqrt{2\tilde{\omega}+q^{2}}/\xi _{N}$, $\xi _{N}=\sqrt{%
D_{N}/2\Delta }$, $q=Q\xi _{N}$ and $f_{S}=\Delta /\zeta _{\omega }$. The
solution describes correctly the condensate Green's function in N provided
the condition $R_{B}>\rho _{N}\xi _{N}$ is fulfilled.

In the limit of a weak PE the energy $F_{N}+F_{B}$ can be written in the form

\begin{equation}
F_{N}+F_{B}=\nu _{N}(2\pi T)D_{N}\sum_{\omega \geqslant 0}\Bigg\{%
\int_{0}^{d_{N}}dx\frac{1}{2}[(\partial _{x}\theta _{N})^{2}+\kappa
_{q}^{2}\theta _{N}^{2}]+\kappa _{B}[1-\cos (\theta _{S}-\theta _{N})]\Bigg\}
\label{9a}
\end{equation}%
where the last term is the boundary free energy \cite{Altland00,Giazotto16}.
Substituting the solution (\ref{9}) into (\ref{9a}), we come to the formula
for $F_{N}+F_{B}$ and $\theta _{S}$ one can easily calculate the

\begin{equation}
F_{N}+F_{B}=\nu _{N}(2\pi T)(D_{N}\kappa _{B}^{2})\sum [\frac{f_{S}^{2}}{%
2\kappa _{q}\tanh (\kappa _{q}d_{N})}+\frac{1}{\kappa _{B}}(1-\cos \theta
_{S})-\frac{f_{S}^{2}}{\kappa _{q}\tanh (\kappa _{q}d_{N})}]  \label{10}
\end{equation}

The first term in the figure brackets is the contribution of the bulk N
region whereas the last term stems from the boundary contribution to the
free energy. The second term is a reduction of the free energy due to the
PE. One can see that the first term gives a positive contribution to the $F$
and decreases with increasing the condensate velocity $V_{S}\sim q$. However
the boundary term (the last one) is twice larger than the first one and
therefore the total contribution of the terms due to condensate current, Eq.(%
\ref{7},\ref{10}), is positive. This means that the condensate current
reduces the free energy.

\textit{Conclusions:} To conclude, we analyzed the free energy for S/N
bilayer in the presence of the condensate current. We have shown that the
bulk of the N film gives a positive contribution $F_{N}(q)$ to the free
energy which decreases with increasing condensate velocity $V\sim q$.
However the contribution of boundary term $F_{B}$ to $F$ is twice larger in
magnitude than $F_{N}(q)$ and is also negative as the contribution $F_{S}$
of the superconductor S. Therefore the total free energy $F$ increases when
condensate moves; this makes the current-carrying state unfavorable.

\textit{Acknowledgements:} We gratefully acknowledge financial support by
the German Research Foundation within the DFG Project ER463/14-1. We thank
also Sebastian Bergeret for pointing out the importance of the boundary term in the free energy functional and Anton Vorontsov and Patric Holmvall for useful comments and discussions.

\setcounter{equation}{0} \setcounter{figure}{0} \setcounter{table}{0}
\makeatletter \renewcommand{\theequation}{S\arabic{equation}} %
\renewcommand{\thefigure}{S\arabic{figure}} \renewcommand{%
\bibnumfmt}[1]{[S#1]} \renewcommand{\citenumfont}[1]{S#1}

\newpage \onecolumngrid \appendix

\section{Supplementary Information}

\setcounter{equation}{0} \setcounter{figure}{0} \setcounter{table}{0}
\renewcommand{\theequation}{S\arabic{equation}} \renewcommand{\thefigure}{S%
\arabic{figure}} \renewcommand{\bibnumfmt}[1]{[S#1]} \renewcommand{%
\citenumfont}[1]{S#1}


\subsection{General Case}

Here we present the evaluation of the free energy $F_{N}$ in the N film. We
integrate once Eq.(7) in the main text
\begin{equation}
-\frac{D_{N}}{2}(\partial _{x}\theta _{N})^{2}+2\omega (1-\cos \theta _{N})+
\frac{D_{N}P_{N}^{2}}{4}[1-\cos (2\theta _{N})]=0  \label{S1}
\end{equation}
and assumed that $d_{N}\gg \xi _{N,\Delta }\equiv \sqrt{D_{N}/2\Delta }$ so
that $\theta _{N}=0$ at $x=d_{N}$. Taking into account Eq.(\ref{S1}), the
energy $F_{N}$ can be written as follows
\begin{equation}
F_{N}=\nu _{N}(2\pi T)D_{N}\sum_{\omega \geqslant 0}\int_{0}^{\infty
}dx(\partial _{x}\theta _{N})^{2}\equiv \nu _{N}(2\pi T)D_{N}\sum_{\omega
\geqslant 0}\tilde{F}_{N,\omega }\text{,}  \label{S2a}
\end{equation}
where $\tilde{\omega}=\omega /\Delta $ and the function $\tilde{F}_{N,\omega
}$ is defined as
\begin{eqnarray}
\tilde{F}_{N,\omega } &=&\int_{0}^{\infty }dx(\partial _{x}\theta
_{N})^{2}=-\int_{0}^{\theta _{N0}}d\theta _{N}(\partial _{x}\theta _{N})=
\label{S3a} \\
&=&-\int_{0}^{\theta _{N0}}d\bar{\theta}_{N}\sin \bar{\theta}_{N}\sqrt{
\kappa _{\omega }^{2}+P^{2}\cos ^{2}\bar{\theta}_{N}}= \\
&=&\frac{P}{2}\left[\sqrt{1+a_{\omega }^{2}}-t_{0}\sqrt{t_{0}^{2}+a_{\omega
}^{2}} +a_{\omega }^{2}\ln \frac{1+\sqrt{1+a_{\omega }^{2}}}{t_{0}+\sqrt{%
t_{0}^{2}+a_{\omega}^{2}}} \right]
\end{eqnarray}
where $t_{0}\equiv \cos \bar{\theta}_{N}\equiv \cos (\theta _{N}/2)|_{x=0}$
and $a_{\omega }^{2}=\kappa _{\omega }^{2}/P^{2}$. The parameter $t_{0}$ is
found from the boundary condition
\begin{equation}
\sqrt{1-t_{0}^{2}}\sqrt{1+(q/\kappa _{\omega }\xi _{N})^{2}t_{0}}=\frac{
\kappa _{BN}}{2\kappa _{\omega }}\left[\Delta (2t^{2}-1)-2\omega t_{0}\sqrt{
1-t_{0}^{2}}\right]\text{,}  \label{S4}
\end{equation}

\subsection{Weak PE}

Consider now a weak PE when the function $\theta _{N}$ is small. In this
case one can obtain a formula for $F_{N}$ for arbitrary thickness $d_{N}$.
At $\theta _{N}\ll 1$, Eq.(2) in the main text can be linearised
\begin{equation}
-\partial _{xx}^{2}\theta _{N}+\kappa _{N}^{2}\theta _{N}=0\text{. }
\label{S5}
\end{equation}
where $\kappa _{N}^{2}=\kappa _{N\omega }^{2}+P^{2}$, $\kappa _{N\omega
}^{2}=2\omega /D_{N}$. The boundary conditions, Eq.(3), have the form
\begin{eqnarray}
\partial _{x}\theta _{N} &=&-\kappa _{B,N}[\sin \theta _{S}-\theta _{N}\cos
\theta _{S}]|_{x=0}\text{, }  \label{S6} \\
\partial _{x}\theta _{N} &=&0|_{x=d_{N}}\text{. }
\end{eqnarray}
where $\kappa _{BN}=2/R_{B}\sigma _{N}$. The solution for Eq.(\ref{S5})
obeying the condition (\ref{S6}) is
\begin{equation}
\theta _{N}(x)=\frac{\kappa _{BN}}{\kappa _{N}}\frac{\cosh (\kappa
_{N}(x-d_{N}))}{\cosh \vartheta _{N}\mathit{D}_{N}}\sin \theta _{S}\text{. }
\label{S7}
\end{equation}
where $\mathit{D}_{N}=\tanh \alpha _{N}+(\kappa _{BN}/\kappa _{N})\tilde{
\omega}/\sqrt{\tilde{\omega}^{2}+1}$, $\alpha _{N}=\kappa _{N}d_{N}$, $\sin
\theta _{S}=1/\sqrt{\tilde{\omega}^{2}+1}$. The energy of the N film, $F_{N}
$, is
\begin{eqnarray}
F_{N} &=&2\nu _{N}(2\pi T)\sum_{\omega \geqslant 0}\int_{0}^{d_{N}}dx\left[%
\frac{ D_{N}}{4}(\partial _{x}\theta _{N})^{2}+\frac{1}{4}(2\omega
+D_{N}P^{2})\theta _{N}^{2} \right]=  \label{S8} \\
&=&2\nu _{N}\frac{D_{N}\kappa _{BN}^{2}}{4}\xi _{N,\Delta }(2\pi
T)\sum_{\omega \geqslant 0}\frac{\tanh \alpha _{N}}{\mathit{D}_{N}^{2}}\frac{
1}{\sqrt{\tilde{\omega}+q^{2}}}\frac{1}{\tilde{\omega}^{2}+1}
\end{eqnarray}
In the limit of a thick N film ($d_{N}\gg \xi _{N,\Delta }$) Eq.(\ref{S6})
acquires the form

\begin{equation}
F_{N}=2\nu _{N}\frac{D_{N}\kappa _{BN}^{2}}{4}\xi _{N,\Delta }(2\pi
T)\sum_{\omega \geqslant 0}\frac{1}{\sqrt{\tilde{\omega}+q^{2}}}\frac{1}{%
\tilde{\omega}^{2}+1}  \label{S9}
\end{equation}

\end{document}